\documentclass{article}

\usepackage{microtype}
\usepackage{graphicx}
\usepackage{subcaption}
\usepackage{booktabs}
\usepackage{hyperref}

\usepackage[accepted]{icml2026}

\usepackage{amsmath}
\usepackage{amssymb}
\usepackage{mathtools}
\usepackage{amsthm} 
\usepackage[capitalize,noabbrev]{cleveref}

\theoremstyle{plain}

\theoremstyle{definition}

\theoremstyle{remark}

\usepackage[textsize=tiny]{todonotes} 
\usepackage{hyperref}
\usepackage{url}
\usepackage{multirow}
\usepackage{booktabs}      
\usepackage{amsfonts}        
\usepackage{nicefrac}        
\usepackage{microtype}      
\usepackage{xcolor}         
\usepackage{makecell}
\usepackage{adjustbox}
\usepackage{pifont}   
\newcommand{\cmark}{\ding{51}} 
\newcommand{\xmark}{\ding{55}} 
\usepackage[table]{xcolor} 
 
\icmltitlerunning{ReGen: Hierarchical Multi-Prompt Representation Generation for Efficient Waveform Diffusion Models}

\begin{document} 

\twocolumn[
  \icmltitle{ReGen: Hierarchical Multi-Prompt Representation Generation for \\Efficient Waveform Diffusion Models}
 
  \icmlsetsymbol{equal}{*}

  \begin{icmlauthorlist}
    \icmlauthor{Sang-Hoon Lee}{1}
    \icmlauthor{Ha-Yeong Choi}{comp} 
  \end{icmlauthorlist}

  \icmlaffiliation{1}{Department of Artificial Intelligence, Ajou University, Suwon, Korea}
  \icmlaffiliation{comp}{KT Corp., Seoul, Korea} 

  \icmlcorrespondingauthor{Sang-Hoon  Lee}{sanghoonlee@ajou.ac.kr} 
  \icmlkeywords{Machine Learning, ICML}

  \vskip 0.3in
]
 
\printAffiliationsAndNotice{} 

\begin{abstract}

Representation alignment (REPA) has been investigated to accelerate diffusion training, but we observe that regularizing intermediate representations in diffusion Transformers (DiT) may implicitly entangle latents and limit generative capacity. 
To address this issue, we propose ReGen, a hierarchical multi-prompt representation generation framework that jointly estimates multiple vector fields for both representations and data within a single diffusion model.
We further introduce generalized flow matching (GFM) to improve the generalization of conditional flow matching (CFM). We validate ReGen on single-stage waveform diffusion models including neural audio codec and Wave-VAE. ReGen significantly improves waveform generation quality from highly compressed latent representations at 12.5 Hz. We also present ReGenVoice, a latent diffusion model (LDM)-based text-to-speech model that achieves strong speech intelligibility (WER) and speaker similarity (SIM) with a small dataset. Moreover, operating the LDM at 6.25 Hz with rich semantic and acoustic latent representation enables efficient training and sampling, requiring only 1 day of training on 4 GPUs and fast inference with an RTF of 0.08. Audio samples are available at \url{https://regenvoice.github.io/demo/}. 
\end{abstract}

\section{Introduction} 
Recently, audio has been getting interest in human-centric artificial intelligence systems, enabling both audio understanding and generation in spoken dialogue systems. To do this, it is essential to compress high-resolution raw waveform signals into low-resolution latent representations via vector quantization (VQ) or variational autoencoders (VAEs), followed by large language models that predict VQ-based discrete audio tokens and diffusion models that generate VAE-based continuous audio vectors. Then, audio decoder converts these representation into an audible waveform. 

While generative adversarial networks (GAN)-based neural audio codecs \cite{zeghidour2021soundstream} are commonly used, they have several drawbacks in extremely low-bitrate scenarios: 1) limited semantic capacity, which reduces speech intelligibility; 2) limited acoustic capacity, which degrades high-frequency details. Basically, semantic distillation with residual vector quantization (RVQ) \cite{zhang2024speechtokenizer, defossez2024moshi} has been widely used to improve the semantic consistency. However, it requires additional quantization and can reduce generative capacity. 

Meanwhile, conditional flow matching (CFM)-based models have gained increasing attention as a new generation paradigm for audio modeling  \cite{lee2025periodwave, liu2025rfwave, welker2025flowdec,yao2025flow2gan}. PeriodWave-Turbo \cite{lee2024accelerating} demonstrated that CFM-based pre-training and adversarial post-training significantly improve the performance and reduce overall training times. StreamFlow \cite{choi2026streamflow} introduces streaming flow matching for streaming DiT-based waveform generation, and analyzes representation alignment (REPA) \cite{yu2025representation} learning in diffusion Transformers (DiT) \cite{peebles2023scalable} for waveform generation, where REPA accelerates training speed and can improve generative capacity via robust semantic alignment. However, recent studies \cite{wang2025repa} report that REPA can suffer from capacity mismatch that decreases the capacity on generative ability due to implicitly entangled latent representations in DiT.    

To address this limitation, we shift our focus from representation alignment to representation generation (ReGen) for robust waveform DiT training. Rather than regularizing intermediate representations to maximize semantic alignment, we explicitly disentangle semantic features from the waveform by jointly estimating multiple vector fields for both representations and data within a single diffusion model. We carefully design a hierarchical DiT structure that progresses from semantics to waveform across layers. Moreover, we introduce a multi-prompting mechanism that hierarchically guides both representation and waveform generation via a masked-infilling strategy. Moreover, we present generalized flow matching (GFM) for robust waveform-level flow matching via a repulsive term in the vector field space.

Specifically, we demonstrate the effectiveness of ReGen and GFM on a low-bitrate neural audio codec (25 Hz, 400 bps) and a VAE (12.5 Hz, 32 dimension) for 24 kHz waveform generation. Furthermore, we propose ReGenVoice, a latent diffusion models (LDM)-based text-to-speech model, performing the diffusion process at 6.25 Hz. With a small-scale dataset and only one day of training on 4 GPUs, the model achieves strong performance in speech intelligibility and speaker similarity. 
The main contributions are as follows: 
\vspace{-0.2cm}
\begin{itemize}
\item We introduce \textbf{ReGen}, a hierarchical representation prompting and generation framework built within a single DiT.
\item We propose \textbf{GFM}, generalized flow matching that improves robustness in waveform-level flow matching training by mitigating zero-collapse.
\item We demonstrate the effectiveness of ReGen and GFM by achieving state-of-the-art performance in speech reconstruction and prompted generation.
\item We introduce an LDM-based TTS model, ReGenVoice to verify the effectiveness of ReGen in generative modeling, achieving strong speech intelligibility and speaker similarity. We will release all source code. 
\end{itemize} \vspace{-0.2cm}

\vspace{-0.2cm}
\section{Neural Waveform Generation with SSL}
In this section, we summarize self-supervised learning (SSL) representation for speech, the semantic distillation for neural audio codec and the representation alignment (REPA) for recent diffusion Transformers (DiT)-based neural waveform generation, and their limitations. \vspace{-0.1cm}
\vspace{-0.1cm}
\subsection{Self-supervised Representation for Speech}
In the speech domain, leveraging SSL representation is a common approach for improving the semantic capacity on highly entangled audio representations. \cite{polyak21_interspeech} first adopted self-supervised discrete representations \cite{hsu2021hubert} to disentangle speech into content, prosody, and speaker identity. NANSY \cite{choi2021neural} used self-supervised continuous representation \cite{baevski2020wav2vec} as a linguistic representation. HierSpeech \cite{lee2022hierspeech} utilized SSL representation as an intermediate representation between text and waveform. Miipher \cite{koizumi2023miipher} used w2v-BERT \cite{chung2021w2v} for speech enhancement. SPEAR-TTS \cite{kharitonov2023speak} and Make-A-Voice \cite{huang2024make} introduce SSL based semantic token and residual vector quantization (RVQ)-based acoustic token in a hierarchical manner. RepCodec \cite{huang2024repcodec} directly used SSL representation as an input and quantized them as a semantic token. X-Codec series \cite{ye2024codecdoesmatterexploring, ye2025llasa} jointly used SSL representation as an input representation with waveforms to improve semantic capacity for low-bitrate neural audio codec.
\vspace{-0.1cm}
\subsection{Semantic Distillation}
While directly using SSL representations as intermediate features or semantic tokens is effective, it often increases inference complexity and may require additional modules during inference. 
As an alternative, several works exploit SSL representations only as supervision during training, transferring high-level semantic information into compact latent representations without explicitly relying on SSL features at inference time. 
SpeechTokenizer \cite{zhang2024speechtokenizer} introduced semantic distillation objectives to improve the linguistic capacity and robustness of low-bitrate neural audio codecs by aligning the first codec latents with SSL representations. Mimi \cite{defossez2024moshi} used a group RVQ approach by disentangling semantic VQ and acoustic RVQ for robust downstream tasks.
 
\vspace{-0.1cm}
\subsection{Representation Alignment (REPA)}
Recently, conditional flow matching (CFM)-based neural waveform generation model have shown promising performance. Among them, StreamFlow \cite{choi2026streamflow} is the only method that adopts DiT for waveform generation, and introduces scale normalization layers for robust waveform-level diffusion. Furthermore, StreamFlow leverages REPA on the intermediate DiT blocks by using an intermediate representation from MMS \cite{pratap2024scaling} as a speech-level semantic representation, leading to faster convergence and improved semantic consistency.
\vspace{-0.1cm}
\subsection{Limitations} Building on StreamFlow, we train a single-stage waveform diffusion neural audio codec from scratch and incorporate REPA using semantic representations. However, we observe that learning meaningful information in extremely low-bitrate settings remains challenging, even with REPA. 
To mitigate this issue, we additionally introduce acoustic REPA using mel-spectrograms, as well as hierarchical REPA (REPA-H) to enhance both semantic and acoustic capacity. While REPA-H enables stable training, REPA alone cannot guarantee consistent speech generation in terms of semantic and acoustic characteristics, due to severe information loss in highly compressed latent representations. 
To address this limitation, we shift our focus from REPA to representation generation (ReGen) for more robust waveform diffusion models. Experimental results for REPA and REPA-H are summarized in Table \ref{ablation1} and the details of ReGen are described in the following subsections.

\begin{figure*}
    \centering \vspace{-0.3cm}
    {\includegraphics[width=0.95\textwidth]{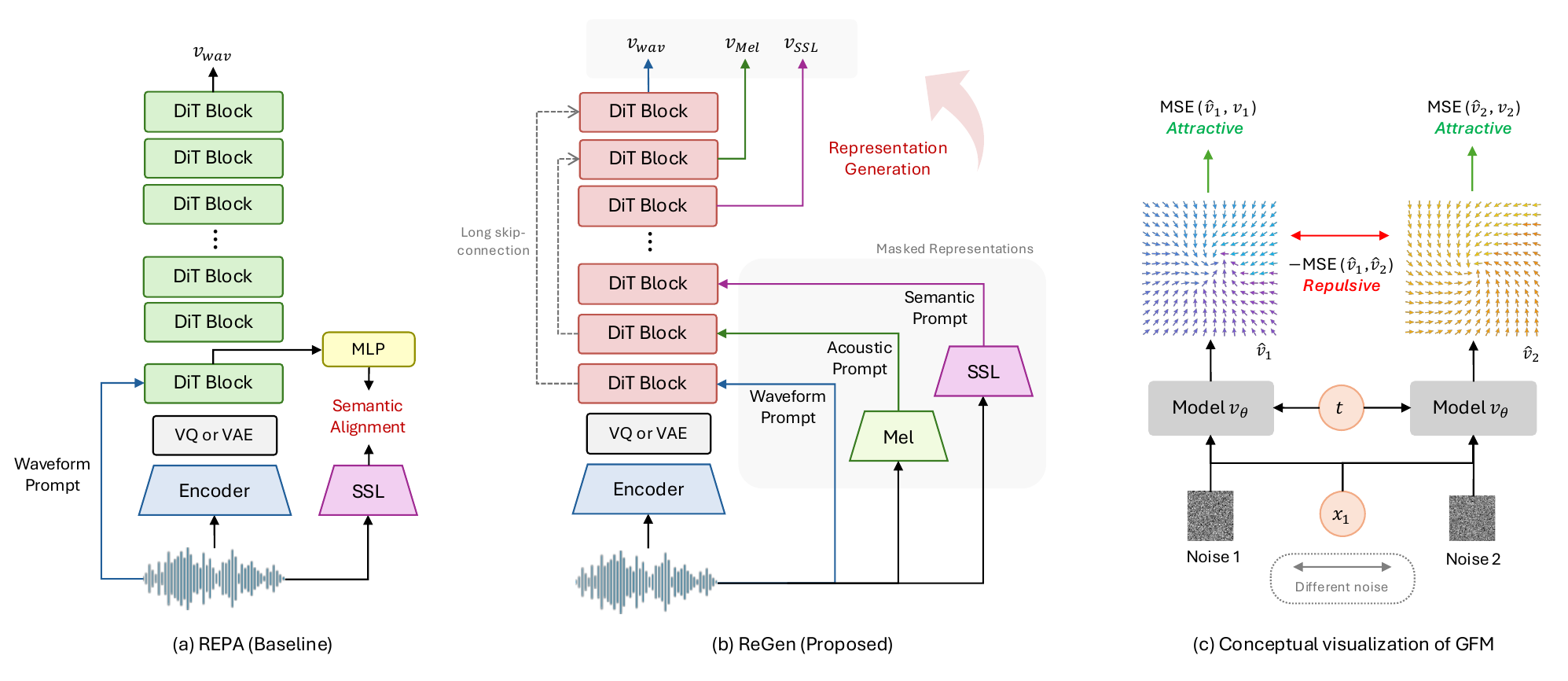}}\vspace{-0.4cm}
    \caption{(a) A baseline that applies REPA to waveform generation, (b) the proposed ReGen architecture with hierarchical multi-prompting, and (c) a conceptual visualization of generalized flow matching (GFM)}
    \label{fig:architecture} \vspace{-0.4cm}
\end{figure*}

\section{ReGen} \vspace{-0.1cm}
While REPA can accelerate the early states of diffusion training, it can implicitly entangle latent representations in DiT, resulting in capacity mismatch that reduces generative capacity. This can result in low-quality waveform, especially limiting high-frequency generation while focusing low-frequency details such as semantics of speech. This phenomenon becomes more pronounced in low-bitrate scenarios, where compressed latent representations provide limited acoustic bandwidth and REPA regularization further biases the model toward overly smoothed reconstructions.
\vspace{-0.1cm}
\subsection{Representation Generation} \vspace{-0.1cm}
To avoid the above problems, we introduce ReGen that jointly estimates the vector field of representation and data and generate together through iterative sampling. Unlike prior approaches that treat representations as fixed conditioning signals, ReGen models the representation as a stochastic variable to be generated, thereby reducing over-reliance on the conditions and explicitly disentangling representation from data. As a result, the generative model can flexibly allocate its capacity to synthesize both high-level semantics and acoustic details, even under low-bitrate settings. 

For robust generation in terms of architecture designs, we disentangle the DiT block by generating from representations to data in a hierarchical manner. Additionally, we leverage multiple representations including semantic (SSL) and acoustic (Mel-spectrogram) representations by estimating multiple vector fields on waveform, Mel-spectrogram, and SSL representation together from high-compressed latent representation of VQ or VAE.  
\vspace{-0.1cm}
\subsection{Hierarchical Multi-Prompt} \vspace{-0.1cm}

ReGen extends the hierarchical multi-prompt design to simultaneously utilize information from different representation levels. Rather than simply providing conditional input, each prompt provides direct information on the generation process at the semantic, acoustic, and waveform levels within each representation space. 
The proposed method predicts independent vector fields for each representation, while simultaneously employing a hierarchical structure in which information from higher-level representations progressively constrains the generation of lower-level representations. 

For each representation level $r\in\{\mathrm{ssl}, \mathrm{mel}, \mathrm{wav}\}$, the conditional flow between the target representation $x_1^r$ and the noise sampled from $x_0^r \sim \mathcal{N}(0,I)$ is defined according to the time $t \sim \mathcal{U}(0,1)$ as follows:
\vspace{-0.1cm}
\begin{equation}
x_t^r = \bigl(1 - (1 - \sigma_{\min}) t\bigr) x_0^r + t x_1^r 
\end{equation} \vspace{-0.1cm}
The corresponding target vector field is given by: \vspace{-0.1cm}
\begin{equation}
    u_t^r = x_1^r-(1 -\sigma_{\min}) x_0^r
\end{equation} \vspace{-0.1cm}
During training, a mask $M$ is applied to all prompts, $\tilde{x}_1^r = M \odot x_1^r$. Then, the network predicts the vector field $\hat{v}_{\theta}^r$ for each level conditioned on masked multi-prompt. This masked multi-prompt infilling strategy \cite{le2024voicebox} enables ReGen to learn robust vector fields. Finally, the masked multi-prompt CFM objective is defined as:\vspace{-0.1cm}
 \begin{equation}
\small
\mathcal{L}_{\text{CFM}}=\sum_{r \in \{\text{ssl},\text{mel},\text{wav}\}} \mathbb{E}_{t, x_0, x_1} \left[ \bigl\| \hat{v}_{\theta}^{r}(x_t^{r}, t, \tilde{x}_1^{r}) - u_t^{r} \bigr\|^2 \odot M \right]
\end{equation}   

\vspace{-0.3cm}
\section{Generalized Flow Matching}  
Waveform generation tasks face significant challenges in regression-based learning due to the multi-modal nature and phase inconsistencies of high-frequency components. Under a standard regression objective, these characteristics force diverse potential waveforms to converge towards a single average solution. This often results in over-smoothing, where the generated waveform lacks detail. When the vector field is trained via regression, different probabilistic trajectories may collapse into a single conditional mean flow. Consequently, the generated waveforms tend to lose high-frequency energy, leading to degraded audio quality.
To mitigate this, \cite{gritsenko2020spectral} was proposed that introduces a repulsive term between generated samples using the generalized energy distance (GED). As discussed in \cite{gritsenko2020spectral}, using standard regression loss without repulsive regularization causes samples to converge to the conditional mean or produces waveforms with distorted phase and energy in high-dimensional space. Inspired by these observations, we extend this principle to conditional flow matching. Specifically, we reformulate sample-space repulsive regularization to fit vector field optimization. We propose generalized flow matching (GFM), which reformulates repulsive regularization in the velocity space, aligning the objective with the optimization domain of CFMs. 

We explicitly control the similarity between predicted velocity vectors to prevent multiple stochastic trajectories from collapsing into a single flow. To ensure training stability in high-dimensional space, we adopt a mechanism where one trajectory serves as a fixed reference, while the other is pushed away from it and simultaneously pulled towards the ground-truth target trajectory. This mitigates the tendency to converge towards an average solution while effectively preventing divergence during the training process.
Formally, consider two flow trajectories $x_t^{(1)}$ and $x_t^{(2)}$ that are coupled with the same target waveform but originate from different initial noise samples $x_0^{(1)}$ and $x_0^{(2)}$. Let $u_t^{(1)}$ and $u_t^{(2)}$ be their respective target vector fields. We define the attractive loss $\mathcal{L}_{\text{att}}$ and the asymmetric repulsive loss $\mathcal{L}_{\text{rep}}$ as: \vspace{-0.1cm} 
\begin{equation}
\small
    \mathcal{L}_{\text{att}} = \frac{1}{2} \left( \| v_\theta(x_t^{(1)}, t,c) - u_t^{(1)} \|^2 + \| v_\theta(x_t^{(2)}, t, c) - u_t^{(2)} \|^2 \right)
\end{equation}
\begin{equation}
    \mathcal{L}_{\text{rep}} = \| v_\theta(x_t^{(1)}, t, c) - \text{sg}(v_\theta( x_t^{(2)}, t, c)) \|^2
\end{equation}
where $v_\theta$ denotes the velocity estimator network, and $\text{sg}(\cdot)$ denotes the stop-gradient operator. The GFM objective is constructed by subtracting the repulsive term from the attractive term, controlled by $\lambda_{\text{neg}}$:\vspace{-0.1cm} 
\begin{equation}
    \mathcal{L}_{\text{GFM}} = \mathcal{L}_{\text{att}} - \lambda_{\text{neg}}\mathcal{L}_{\text{rep}}
\end{equation}  \vspace{-0.1cm}
We apply GFM objectives independently in waveform, Mel-spectrogram, and SSL representation spaces. The final training objective is:\vspace{-0.1cm}
\begin{equation}
\mathcal{L}
= \mathcal{L}_{\mathrm{GFM}}^{\mathrm{wav}} + \lambda_{regen} \left( \mathcal{L}_{\mathrm{GFM}}^{\mathrm{Mel}}
+ \mathcal{L}_{\mathrm{GFM}}^{\mathrm{SSL}} \right).
\end{equation}  \vspace{-0.1cm}

\vspace{-0.8cm} 
\section{Efficient Waveform Diffusion Transformers} \vspace{-0.1cm}
\paragraph{Wave-DiT}
For an efficient waveform diffusion Transformers (Wave-DiT), we utilize a linear-reshape Transformation and Scale-DiT of StreamFlow \cite{choi2026streamflow} to operate diffusion at 50 Hz for 24,000 Hz waveform signals directly. Also, we modify Scale-DiT by replacing AdaLN with AdaLN-SOLA \cite{hai2024ezaudio} for parameter-efficient shared time adaptation. We add hierarchical multi-prompt embedding layer for each representation and U-Net like long-skip connection (Add) for disentangled representation generation as illustrated in Figure \ref{fig:architecture}. Specifically, Wave-DiT consists of Wave-DiT blocks of [3,3,6,3,3] with 1024 dimensions at 50 Hz. To demonstrate the powerful generative performance of ReGen, we train both neural audio codec and VAE encoder using Wave-DiT decoder.     
\vspace{-0.1cm}
\paragraph{ReGenTokenizer}
For neural audio codec, we also use a linear-reshape transformation for input waveform (from 24,000 Hz to 50 Hz), followed by eight layers of causal Transformers using RoPE, context window of 4s, and dimension of 1024. Then, we simply downsample and apply finite scalar quantization (FSQ) using codebook size of 65536 (FSQ of [4,4,4,4,4,4,4]) at 25 Hz (400bps). The upsampled outputs are fed to Wave-DiT as a condition for waveform generation at 50 Hz. 
\vspace{-0.1cm}

\paragraph{ReGenVAE}
We use the same linear-reshape transformation and eight layers of causal Transformers. Then, we simply downsample and apply weak KL regularization ($\lambda_{\text{kl}}$ of 1e-5) on the highly compressed latent representations (32 dimensions at 12.5 Hz). The upsampled latent representations are used as a condition for Wave-DiT.

\begin{table*}[th!]
 \caption{Objective evaluation results for reconstruction using LibriSpeech benchmark dataset. For reference, we cite the reported performance of models from X-codec2 \citep{ye2025llasa}, and use the same evaluation methods used in X-codec2.}  \label{table1:librispeech} \vspace{-0.2cm}
\centering
  \resizebox{0.98\textwidth}{!}{
\begin{tabular}{l|cccc|c|c|c|cc|c|c}
\toprule
Method& Hz & TPS & $N_q$ & Bitrate&Stream.   & WER ($\downarrow$)& STOI ($\uparrow$) & \makecell{PESQ\\ -WB ($\uparrow$)} & \makecell{PESQ\\-NB ($\uparrow$)} &SPK-SIM ($\uparrow$) & UTMOS ($\uparrow$) \\
    \midrule
    GT & 16k & -& -& - & -& 1.96&1.00&4.64&4.55&1.00&4.09   \\
\midrule%
SpeechTokenizer \citep{zhang2024speechtokenizer}  &16k & 100 & 2 &1000  & \cmark& 3.92   & 0.77  & 1.25    & 1.59 & 0.36 & 2.28 \\
\midrule 
BigCodec \citep{xin2024bigcodec}  &16k & 80 & 1 &1040 & \xmark& 2.76&0.93&2.68&3.27&0.84&4.11 \\
X-codec2 \citep{ye2025llasa} &16k & 50 & 1 &800 & \xmark& 2.47&0.92&2.43&3.04&0.82&4.13 \\
 StableCodec\citep{parker2025scaling}  &16k & 50 & 2 & 700 & \xmark &5.12&0.91&2.24&2.91&0.62&4.23 \\
\midrule 
EnCodec \citep{dfossez2023high} &24k & 600 & 8 &6000 & \cmark & 2.15   & 0.94  & 2.77    & 3.18 & 0.89 & 3.09  \\
EnCodec \citep{dfossez2023high} &24k & 150 & 2 &1500  & \cmark & 4.90   & 0.85  & 1.56    & 1.94 & 0.60 & 1.58  \\
\midrule
WavTokenizer \citep{ji2025wavtokenizer} &24k & 75 & 1 &900 & \xmark & 3.98&0.90&2.13&2.63&0.65&3.79  \\
WavTokenizer\citep{ji2025wavtokenizer} &24k & 40 & 1 & 480& \xmark & 11.20&0.85&1.62&2.06&0.48&3.57  \\
\midrule
Mimi \citep{defossez2024moshi}&24k & 100 & 8  & 1100 & \cmark& 2.92&0.90& 2.27&2.80&0.73&3.63  \\ 
 \midrule
\textbf{ReGenTokenizer-LibriTTS} &24k & 25 & 1 & 400  & \cmark  & 2.83&0.84&1.39& 1.65 &0.67&3.99 \\
\textbf{ReGenTokenizer-Emilia} &24k & 25 & 1 & 400  & \cmark  & 2.70&0.86&1.52& 1.85 &0.74&3.77 \\
\midrule
\multicolumn{10}{l}{\textbf{\textit{VAE}}} \\  
\midrule
\textbf{ReGenVAE-LibriTTS} &24k & 12.5 & - & - & \cmark & 2.05& 0.94&2.90&3.38&0.84&4.21  \\
\textbf{ReGenVAE-Emilia} &24k & 12.5 & - & - & \cmark & 2.11&0.95&2.99&3.45&0.89&4.16 \\
\bottomrule
\end{tabular}
  }  \vspace{-0.2cm}
\end{table*}

\begin{table*}[th!]
\caption{Objective evaluation results for reconstruction using LibriTTS Benchmark Dataset.}  \label{table2:libritts} \vspace{-0.2cm}
\centering
\resizebox{0.98\textwidth}{!}{
\begin{tabular}{l|ccc|c|cc|ccccc|cc}
\toprule
Method& Hz & TPS & $N_q$&Stream.   & \makecell{CER\\($\downarrow$)} &\makecell{WER\\($\downarrow$)}&  \makecell{M-STFT\\($\downarrow$)}& \makecell{PESQ\\($\uparrow$)} & \makecell{Period.\\($\downarrow$)}&  \makecell{V/UV\\($\uparrow$)}& \makecell{Pitch\\($\downarrow$)} &\makecell{UTMOS\\($\uparrow$)} &\makecell{MOS\\($\uparrow$)}  \\
\midrule
GT & 24k& -& -& -&  1.12&3.06  &-&- & -& -&-& 3.862 & 3.99$\pm$0.02\\
\midrule
SpeechTokenizer \citep{zhang2024speechtokenizer} &16k & 400 & 8  & \cmark&2.06&4.46&-&2.468&0.129&0.934&41.631&3.585  & 3.68$\pm$0.03  \\
\midrule 
BigCodec \citep{xin2024bigcodec} &16k & 80 & 1  & \xmark& 2.61& 5.44& -&2.607&0.145&0.927&33.757&3.889 & 3.77$\pm$0.03  \\
X-codec2 \citep{ye2025llasa} &16k & 50 & 1  & \xmark& 2.45& 5.20 & -&2.259&0.230&0.857&53.052&3.869 & 3.71$\pm$0.03  \\
StableCodec \citep{parker2025scaling} &16k & 50 & 2  & \xmark & 6.85& 12.42& - & 2.072&0.184&0.899&32.002&4.134 & 3.76$\pm$0.03  \\
StableCodec \citep{parker2025scaling} &16k & 25 & 1  & \xmark & 8.62& 16.19& - & 1.893&0.194&0.892&48.680&4.131 & 3.75$\pm$0.03  \\
\midrule 
EnCodec \citep{dfossez2023high} &24k & 600 & 8 & \cmark & 1.19&3.55&1.163&2.771&0.113&0.941&32.147&2.969 & 3.68$\pm$0.03  \\
DAC \citep{kumar2023high} &24k & 600 & 8 & \xmark & 1.09&2.91&1.012 &3.505 &0.075&0.962&22.290&3.546 & 3.82$\pm$0.03  \\
\midrule
Mimi \citep{defossez2024moshi} &24k & 100 & 8   & \cmark& 3.07& 6.91&  1.352&2.266&0.165&0.910&50.686&3.506 & 3.82$\pm$0.03  \\
\midrule
WavTokenizer \citep{ji2025wavtokenizer} &24k & 75 & 1 & \xmark & 4.87& 9.42 & 1.220&2.247&0.157&0.918&41.123&3.821 & 3.86$\pm$0.03  \\
WavTokenizer \citep{ji2025wavtokenizer}  &24k & 40 & 1 & \xmark & 13.83& 23.53 & 1.627&1.604&0.196&0.891&69.903&3.467 & 3.75$\pm$0.03  \\
\midrule
\textbf{ReGenTokenizer-LibriTTS} & 24k & 25 & 1 & \cmark & 1.92&4.43&1.504&1.496&0.184&0.903&94.513&3.955 & 3.88$\pm$0.03  \\
\textbf{ReGenTokenizer-Emilia} & 24k & 25 & 1 & \cmark & 2.23&4.84&1.454&1.606&0.178&0.907&83.194&3.785& 3.82$\pm$0.03  \\
\midrule
\multicolumn{10}{l}{\textbf{\textit{VAE}}} \\  
\midrule
\textbf{ReGenVAE-LibriTTS } & 24k & 12.5 & - & \cmark & 0.94&2.88&1.087&3.058&0.110&0.943&28.299&4.096 & 3.85$\pm$0.03  \\
\textbf{ReGenVAE-Emilia } & 24k & 12.5 & - & \cmark & 1.11&3.18&1.111&2.902&0.115&0.942&30.135&3.956& 3.87$\pm$0.03  \\

\bottomrule
\end{tabular}
}  \vspace{-0.4cm}
\end{table*}
\vspace{-0.1cm}
\paragraph{Acceleration}
Following PeriodWave-Turbo \cite{lee2024accelerating}, we apply adversarial post-training for accelerating the sampling speed by fixing the sampling steps of 4. We employ multi-period discriminator (MPD) \cite{kong2020hifi}, multi-scale STFT discriminator (MS-STFTD) \cite{dfossez2023high}, and multi-scale sub-band constant-Q transform (MS-SB-CQTD) \cite{gu2024multi}. We carefully use multi-scale STFT losses by warming up the weights to maximum 1, where high-weight of STFT losses cause metallic sounds. 

\vspace{-0.1cm}
\section{ReGenVoice}
To verify the effectiveness of ReGen in generative modeling, we train a LDM text-to-speech model, ReGenVoice by using latent representation of ReGenVAE. ReGenVoice consists of causal text encoder, implicit latent aligner, and the same structure of Wave-DiT using down/up-sampling layers. 

\vspace{-0.1cm}
\paragraph{Causal Text Encoder}
For streaming text-to-speech, we utilize six layers of causal Transformer using RoPE, context window of 128 text tokens, and dimension of 1024.  

\vspace{-0.1cm}
\paragraph{Implicit Latent Aligner} 
Inspired by \cite{zhu2025zipvoice}, we use average downsampling for text-latent alignment due to the compressed latents at 12.5 Hz. The output of text encoder are proportionally downsampled and fed to LDM as a condition.

\vspace{-0.1cm}
\paragraph{Efficient Latent Diffusion at 6.25 Hz}
We use U-net like Wave-DiT using down/up-sampling layers for efficient latent diffusion. Operating latent diffusion at 6.25 Hz significantly improves the efficiency and also shows the robust generation performance thanks to high-quality latent representation of ReGenVAE. We train the model using masked-infilling strategy on latent domain and apply GFM objectives for better generalization.

\section{Experiment and result}
\subsection{Experimental Setup}
\paragraph{Dataset} We utilize LibriTTS \cite{zen19_interspeech} for high-quality dataset (0.5k hours) and Emilia \cite{he2024emilia} for high-diversity multi-lingual dataset (100k hours). Both are with a sampling rate of 24,000 Hz, but Emilia is encoded by MP3 so it contains high-frequency distortion. For LDM, we use LibriTTS and Emilia-en subset (40k hours).  
\vspace{-0.5cm}

\paragraph{Training} We pre-train ReGenTokenizer and ReGenVAE with a learning rate of 1$\times$10$^{-4}$ using AdamW optimizer \cite{loshchilov2018decoupled}, batch size of 256 for 1M steps on eight NVIDIA H100 GPUs. We utilize sliced window training by randomly segmenting the waveform signal by 96,000 frames (4s). Then, we fine-tune the four-step model with a learning rate of 2$\times$10$^{-5}$, batch size of 128 for 0.5M steps on eight NVIDIA H100 GPUs. During pre-training, we use masked-infilling strategy by dropping prompt and condition (0.3 and 0.2 respectively). During adversarial post-training, we only drop the prompt by 0.3. For ReGenVoice, we train the model with a learning rate of 1$\times$10$^{-4}$, batch size of 256 for 1M steps on four NVIDIA H100 GPUs using 2-30s latent representations, and drop prompt and condition (0.3 and 0.2 respectively). 

\begin{table*}[th!]
\caption{Objective evaluation results for cross-sentence prompted reconstruction using Seed-en benchmark dataset. Seed-en provides prompt speech samples corresponding to target speech. For reference we cite the reported performance of models from TaDiCodec \cite{wang2025tadicodec}. While previous diffusion models consist of two or three stage, ReGenTokenizer and ReGenVAE consist of only a single decoder for waveform generation. GAN, SPK, Mel-CFM, and Voc. denotes GAN-based waveform generator, speaker encoder, CFM-based Mel-spectrogram decoder, and Mel-to-waveform vocoder, respectively.}  \label{table3:prompted_reconstruction} \vspace{-0.2cm}
\centering
\resizebox{0.98\textwidth}{!}{
\begin{tabular}{l|l|l|ccc|cc|cc|c}
\toprule
\# Stage & Decoder &Method& Hz & TPS & $N_q$&Stream.&Prompt & \makecell{WER\\($\downarrow$)} &\makecell{SPK-SIM\\($\uparrow$)}  &\makecell{UTMOS\\($\uparrow$)} \\
\midrule
Single& GAN&EnCodec \citep{dfossez2023high}  &24k & 150 & 2 & \cmark & \xmark & 5.36&0.48&1.54  \\
&&DAC (RVQ) \citep{kumar2023high}  &24k & 75 & 3 & \xmark& \xmark & 20.08&0.39&1.75  \\
&&DAC (VQ) \citep{kumar2023high} &24k & 75 & 1 & \xmark & \xmark& 12.74&0.45&2.08 \\
&&SpeechTokenizer \citep{zhang2024speechtokenizer}  &16k & 100 & 2  & \cmark& \xmark& 7.98&0.46&2.47\\
&&Mimi  \citep{defossez2024moshi}  &24k & 100 & 8   & \cmark& \xmark& 3.99&0.57&3.21 \\
&&X-codec2  \citep{ye2025llasa}&16k & 50 & 1  & \xmark& \xmark& 2.63&0.62&3.68  \\
&&WavTokenizer \citep{ji2025wavtokenizer}  &24k & 75 & 1 & \xmark& \xmark& 6.65&0.48&3.36\\
& &BigCodec \citep{xin2024bigcodec}&16k & 80 & 1  & \xmark& \xmark& 3.25&0.61&3.59  \\
&&StableCodec \citep{parker2025scaling} &16k & 25 & 1  & \xmark & \xmark& 11.08&0.41&3.87 \\
\midrule
Two & SPK, GAN &BiCodec \cite{wang2025spark}&16k&50&1&\xmark &\cmark&3.05&0.61&3.68 \\
Three & SPK, Mel-CFM, GAN-Voc. & FireRedTTS \cite{guo2024fireredtts} & 48k&25&1&\cmark&\cmark&3.35&0.59&3.40\\
Three &SPK, Mel-CFM, GAN-Voc. & CosyVoice \cite{du2024cosyvoice1} & 24k&25&1&\cmark&\cmark&5.63&0.47&3.65\\
Three & SPK, Mel-CFM, GAN-Voc. & CosyVoice 2  \cite{du2024cosyvoice} & 24k&25&1&\cmark&\cmark&4.10&0.68&3.65\\
\midrule
Three & LDM, MelVAE, GAN-Voc. & SemantiCodec \cite{liu2024semanticodec} & 16k&50&2&\xmark& \xmark &5.11&0.49&2.83\\
Two & Mel-CFM, GAN-Voc. & Vevo \cite{zhang2025vevo} & 24k&50&1&\xmark&\cmark&3.04&0.53&3.50\\
Two & Mel-CFM, GAN-Voc. & TaDiCodec \cite{wang2025tadicodec} & 24k&12.5&1&\xmark&\cmark&2.57&0.69&3.58\\
Two & Mel-CFM, GAN-Voc. & TaDiCodec \cite{wang2025tadicodec} & 24k&6.25&1&\xmark&\cmark&2.73&0.69&3.73\\
\midrule
Single &  Wave-CFM (+GAN) & \textbf{ReGenTokenizer-LibriTTS (ours)} & 24k & 25 & 1 & \cmark& \cmark & 3.37&0.61&3.73  \\
Single & Wave-CFM (+GAN) & \textbf{ReGenTokenizer-Emilia (ours)} & 24k & 25 & 1 & \cmark& \cmark & 3.58&0.71&3.46  \\
\midrule
\multicolumn{10}{l}{\textbf{\textit{VAE}}} \\  
\midrule
Single & Wave-CFM (+GAN) & \textbf{ReGenVAE-LibriTTS (ours)} & 24k & 12.5 & - & \cmark& \cmark & \textbf{2.03}&\textbf{0.72}& \textbf{3.87}  \\
Single & Wave-CFM (+GAN) & \textbf{ReGenVAE-Emilia (ours)} & 24k & 12.5 & - & \cmark& \cmark & \textbf{2.09}&\textbf{0.73}& \textbf{3.73}  \\
\bottomrule
\end{tabular}
}  \vspace{-0.4cm}
\end{table*}

\subsection{Reconstruction}
We conduct two reconstruction experiments on LibriSpeech benchmark dataset (16,000 Hz) from X-codec2 \cite{ye2025llasa} in Table \ref{table1:librispeech} and LibriTTS benchmark dataset (24,000 Hz) from BigVGAN \cite{lee2023bigvgan} in Table \ref{table2:libritts}. ReGenTokenizer successfully demonstrates the effectiveness of ReGen by achieving much better speech intelligibility and speaker similarity compared to other low-bitrate models. Furthermore, we demonstrate the data scalability by training the models with Emilia dataset that shows better speaker similarity. However, Emilia dataset contains a lot of noise and distortion due to MP3 encoding, resulting in high-frequency distortion and low-quality waveform generation.

Additionally, ReGenVAE further demonstrates efficient latent compression with only 32 dimensions at 12.5 Hz by achieving promising speech intelligibility in terms of CER and WER. We also observe that VAE models can enhance audio quality without audio prompts, and prompted generation can mimic the audio quality and style of speech prompts in terms of UTMOS and SPK-SIM as shown in Table \ref{table3:prompted_reconstruction}.

\subsection{Prompted Reconstruction}
Following TaDiCodec \cite{wang2025tadicodec}, we conducted prompted reconstruction using Seed-en benchmark dataset. Table \ref{table3:prompted_reconstruction} also shows the superiority of ReGen in low-bitrate scenarios, where a single-stage ReGen has powerful generative performance through multi-prompting from semantic, acoustic representation, and data. Previous acoustic neural audio codecs including EnCodec, DAC, SpeechTokenizer, Mimi, X-codec2, WavTokenizer, BigCodec, and StableCodec have weak speaker similarity due to lack of style adaptation performance. Semantic speech tokenizer models including BiCodec, FireRedTTS, CosyVoice series require additional speaker embedding model, Mel-spectrogram decoder, and neural vocoder. Furthermore, these models separately train their tokenizer, leading to difficulty in training. 

Meanwhile, TaDiCodec introduced novel frameworks for speech-text foundation models, showing high-quality generation even in extremely low-bitrate scenarios. We thought that ReGen framework can be jointly used with text-aware tokenization for a single-stage decoding. Specifically, ReGenTokenizer shows the best speaker similarity compared to other neural audio codecs, and ReGenVAE has better performance than all models including GT samples (WER: 2.14, SIM: 0.73) by showing the strong generative performance. Also, we hope to highlight that our models are trained with sliced window training (maximum 4s) and short prompts (maximum 1.2s) but show the robust prompting and generation performance (over 30s) as indicated in Table \ref{table3:prompted_reconstruction}. Furthermore, this facilitates the streaming generation using self-generated prompts.       

\begin{table*}[th!]
\caption{Ablation Study. Each model is trained for 0.5M steps, relatively small training steps. Unlike base model (No REPA) and REPA models, ReGen approaches show significantly faster training speed and significantly better performance even with small training steps. ReGen-U* requires smaller weight of representation generation ($\lambda_{\text{regen}}=0.01$) for robust waveform CFM training.}  \label{ablation1}  \vspace{-0.2cm}
\centering
  \resizebox{0.8\textwidth}{!}{
\begin{tabular}{l|cc|ccccc|cc}
\toprule
Method &   \makecell{CER\\($\downarrow$)}&  \makecell{WER\\($\downarrow$)}&  \makecell{M-STFT\\($\downarrow$)} &  \makecell{PESQ\\($\uparrow$)} &  \makecell{Period.\\($\downarrow$)}&  \makecell{V/UV\\($\uparrow$)}&  \makecell{Pitch\\($\downarrow$)} &  \makecell{SPK-SIM\\($\uparrow$)}& \makecell{UTMOS\\($\uparrow$)}\\
  \midrule
 No REPA & 116.88& 136.57 & 2.638& 1.115& 0.236&0.872&126.415&0.08&1.761 \\
 \midrule
\multicolumn{10}{c}{\textbf{\textit{+ Representation Alignment (REPA) }}} \\  
\midrule
 REPA (SSL) & 90.44&130.64&2.536&1.119&0.255&0.846&230.819&0.12&2.159\\
 REPA (Mel) &8.29&15.10&2.060&1.374&0.198&0.892&77.773&0.32&3.090 \\
 REPA-H (SSL, Mel) &9.71&16.53&1.976&1.448&0.189&0.901&70.973&0.38&3.294 \\
  \midrule
\multicolumn{10}{c}{\textbf{\textit{+ Unified Representation Generation (ReGen-U)}}} \\  
\midrule
  ReGen-U (SSL, Mel) & 63.57&89.94&5.412&1.030&0.262&0.808&65.761&0.21&1.338\\
  ReGen-U* (SSL, Mel) & 11.70&19.31&2.152&1.192&0.214&0.879&108.35&0.43&2.114 \\
\midrule
\multicolumn{10}{c}{\textbf{\textit{+ Hierarchical Representation Generation (ReGen-H) }}} \\  
\midrule
  ReGen-H (SSL) & 7.73&15.04&1.985&1.263&0.216&0.879&131.818&0.54&2.862\\
  ReGen-H (Mel) &8.00&14.84& 2.243&1.323&0.199&0.888&76.908&0.35&2.508\\
  ReGen-H (SSL, Mel) &6.36&11.99&1.886&1.452&0.178&0.904&62.536&0.54&3.210 \\
  \midrule
\multicolumn{10}{c}{\textbf{\textit{+ Generalized Flow Matching }}} \\  
\midrule
  ReGenTokenizer & 6.51&12.59&1.795&1.500&0.182&0.904&57.471&0.59&3.268\\
    \midrule
\multicolumn{10}{c}{\textbf{\textit{+ Auxiliary Decoder (Cosyvoice2 token distillation on FSQ)}}} \\  
\midrule
  ReGenTokenizer & 3.09&6.01&1.943&1.315&0.195&0.890&118.072&0.64&2.995\\
\midrule
\multicolumn{10}{c}{\textbf{\textit{+ Replacing FSQ with VAE (12.5 Hz, 32 dim.) } - Auxiliary Decoder}} \\  
\midrule
  ReGenVAE & 1.35& 3.63&1.640&2.063&0.139&0.928&38.953&0.72&3.375 \\
\midrule

\multicolumn{10}{c}{\textbf{\textit{+ Adversarial Post-training}}} \\  
\midrule

ReGenTokenizer & 1.92&4.43&1.504&1.496&0.184&0.903&94.513&0.75&3.955  \\

ReGenVAE & 0.94&2.88&1.087&3.058&0.110&0.943&28.299&0.87&4.096  \\

\bottomrule
\end{tabular}
}\vspace{-0.3cm}
\end{table*}

\begin{table*}[th!]
\caption{Hyper-parameter Search for Generalized Flow Matching}  \label{GFM} \vspace{-0.1cm}
\centering
  \resizebox{0.8\textwidth}{!}{
\begin{tabular}{l|cc|cc|ccccc|cc}
\toprule
Method & $\lambda_{\text{regen}}$ & $\lambda_{\text{neg}}$ & \makecell{CER\\($\downarrow$)}&  \makecell{WER\\($\downarrow$)}&  \makecell{M-STFT\\($\downarrow$)} &  \makecell{PESQ\\($\uparrow$)} &  \makecell{Period.\\($\downarrow$)}&  \makecell{V/UV\\($\uparrow$)}&  \makecell{Pitch\\($\downarrow$)} &  \makecell{SPK-SIM\\($\uparrow$)}& \makecell{UTMOS\\($\uparrow$)}\\
\midrule
ReGen-H & 0.1 & - &6.36&11.99&1.886&1.452&0.178&0.904&62.536&0.54&3.210 \\
\midrule
 & 0.1 & 0.001 & 6.39&13.01&1.814&1.477&0.172&0.911&49.318&0.58&3.254\\
 & 0.1 & 0.005 &7.00&12.28&1.827&1.476&0.172&0.909&47.158&0.58&3.265 \\
  ReGen-H + GFM & 0.1 & 0.01 & 6.51&12.59&1.795&1.500&0.182&0.904&57.471&0.59&3.268\\
& 0.1 & 0.05 & 6.22&12.00&1.951&1.430&0.178&0.905&55.362&0.57&3.269\\
& 0.1 & 0.1 &7.77&14.93&2.059&1.377&0.190&0.896&74.186&0.57&3.072\\
\midrule
Ablation ($\lambda_{regen}$)& 1 & 0.01 &7.43&12.76&2.115&1.361&0.184&0.893&74.832&0.56&2.266\\
& 0.01 & 0.01 & 9.11&16.50&1.897&1.319&0.209&0.881&129.572&0.56&2.955  \\
\bottomrule
\end{tabular}
} \vspace{-0.5cm}
\end{table*}
\vspace{-0.1cm}
\subsection{Ablation Study}
Due to a lack of previous study for waveform-level diffusion neural audio codec, we thoroughly conducted an ablation study for the optimization of diffusion neural audio codec and REPA, ReGen, GFM, and GAN in Table \ref{ablation1}.
\vspace{-0.2cm}
\paragraph{REPA} In extremely low-bitrate scenarios, it is difficult to train the waveform-level diffusion neural audio codec due to the lack of useful latent information on their codebooks without REPA. We found that waveform generation requires more acoustic representations in that Mel-spectrogram distillation on DiT blocks can make the model generate audible waveform signals. Furthermore, we adopt hierarchical REPA (REPA-H) by distilling SSL representation and acoustic representation hierarchically, resulting in better optimization even with the same architecture.

\vspace{-0.2cm}

\paragraph{ReGen} We compare ReGen by unified prompting/generation at the same layer (ReGen-U) or hierarchical prompting/generation (ReGen-H). Disentangling prompt and generation can improve the robustness of ReGen so we adopt ReGen-H as the default settings. While REPA still shows a lack of speaker similarity, ReGen models significantly improve the speaker similarity and all performance compared to REPA. We found that SSL representation can guide the speech-level semantics, and Mel-spectrogram can guide the acoustics of the waveform. 
\vspace{-0.2cm}

\paragraph{GFM} GFM improves the performance compared to the original CFM. We also conducted the hyper-parameter search for GFM at Table \ref{GFM}. While we conducted this hyper-parameter search by using global weights for each representation, we see that further optimization can be possible by analyzing the proper weights for each representation.

\vspace{-0.1cm}

\paragraph{Adversarial Post-training} With only a few steps, adversarial post-training simply accelerates the models and improves the entire performance by directly generating waveform from noise. This indicates that CFM pre-training followed by adversarial post-training is the natural and essential choice to efficiently train waveform generation models.

\begin{table}[t]
    \centering
    \caption{Zero-shot TTS results on Seed-en Benchmark} \vspace{-0.2cm}
    \label{tab:tts_seed_benchmark}
    \resizebox{1\linewidth}{!}{
    \begin{tabular}{l|cc|cc}
        \toprule 
        Model  & Params & Data & \textbf{WER} $\downarrow$ & \textbf{SIM} $\uparrow$ \\
        \midrule
        GT & - & -  & 2.14 & 0.73\\
        \midrule
        SparkTTS \cite{wang2025spark}  & 0.5B & 100k & 1.98 & 0.58\\  
        TadiCodec-AR  \cite{wang2025tadicodec}   & 4B & 100k & 2.28 & 0.65 \\
        \midrule
        F5-TTS \cite{chen2024f5} & 0.3B & 100k & 2.00 & 0.67  \\
        ZipVoice \cite{zhu2025zipvoice}& 0.1B & 100k & 1.70 & 0.69  \\
        CosyVoice2 \cite{du2024cosyvoice} & 0.5B & 166k & 2.57 & 0.65  \\
        CosyVoice3-RL \cite{du2025cosyvoice3}  & 0.5B & - & 1.68 & 0.69  \\
        \midrule
        VibeVoice \cite{peng2025vibevoice}  & 1.5B & - & 3.04 & 0.68  \\
        VibeVoice-RT \cite{peng2025vibevoice} & 0.5B & - & 2.05 & 0.63  \\
        VoxCPM-Emilia \cite{zhou2025voxcpm}  & 0.6B & 100k & 2.34 & 0.68 \\
        \midrule
        ReGenVoice-S  & 0.3B & 0.5k & 2.04& 0.65 \\ 
        ReGenVoice-S  & 0.3B & 40k & 2.21& 0.68  \\ 
        ReGenVoice  & 0.5B & 0.5k & \textbf{1.46}& 0.64 \\ 
        ReGenVoice& 0.5B & 40k & 1.62& \textbf{0.70} \\ 
        \bottomrule
    \end{tabular}
    } \vspace{-0.4cm}
\end{table}

\begin{table}[th!]
    \centering
    \caption{Subjective evaluation and RTF comparison} \vspace{-0.2cm}
    \label{tab:tts_mos}
    \resizebox{0.98\linewidth}{!}{
    \begin{tabular}{l|cc|c}
        \toprule  
        Model  & MOS $\uparrow$ & SMOS $\uparrow$ & RTF  $\downarrow$\\
        \midrule
        GT &   3.978$\pm$0.02 &  3.756$\pm$0.02 & -\\
        \midrule
        CosyVoice2 & 4.033$\pm$0.02 &  3.754$\pm$0.02 & 1.05  \\
        CosyVoice3-RL (vLLM) & 3.867$\pm$0.02 & 3.606$\pm$0.02 & 0.81  \\
        \midrule
        ReGenVoice (25 NFE)& 4.029$\pm$0.02 & 3.762$\pm$0.02 & 0.08 \\ 
        \bottomrule
    \end{tabular}
    }\vspace{-0.5cm}
\end{table}
\subsection{Text-to-Speech}
\paragraph{Experiment} We conducted zero-shot TTS experiments on Seed-en benchmark dataset. We train our models with LibriTTS and Emilia-en dataset using 0.3B/0.5B parameters to verify the scalability. We compared our models with recent LLM-based TTS models, CFM-based Mel-spectrogram generation models, and latent diffusion models. Specifically, our ReGenVoice employs the latent diffusion at 6.25 Hz using down/upsampling layers for latents of ReGenVAE, resulting in fast sampling speed without any optimization technique such as vLLM, Triton, and FlashAttention. 

\vspace{-0.1cm}
\paragraph{Result} Table \ref{tab:tts_seed_benchmark} shows the powerful generative performance of ReGen frameworks by achieving lower WER and higher speaker Similarity compared to others. Even with a small-scale dataset, our models achieve comparable performance in terms of speaker similarity. Specifically, it is worth noting that ReGenVAE was trained with a small-scale of STFT loss that reduces the performance on objective evaluation including WER and SIM. However, it significantly improves the perceptual quality in human evaluation as indicated in Table \ref{tab:tts_mos}, where some models generate noisy metallic sounds due to overuse of STFT losses for high objective performance. When using high STFT loss, ReGenVoice achieved a speaker similarity of 0.72, but it may contain metallic noise in some cases. 

\begin{table}[th!]
    \centering
    \caption{Ablation study for ReGenTTS} \vspace{-0.2cm}
    \label{tab:tts_ablation_seed_benchmark}
    \resizebox{0.98\linewidth}{!}{
    \begin{tabular}{l|c|ccc}
        \toprule

        Ablation & Method &  \textbf{WER} $\downarrow$ & \textbf{SIM} $\uparrow$ & \textbf{UTMOS} $\uparrow$ \\
        \midrule
        ReGenVoice & -& 1.46& 0.64 & 4.07 \\ 
           \midrule
        Text Encoder &w/o Encoder& 1.38& 0.61 & 4.09 \\ 
                 & ConvNeXt V2& 2.13& 0.56 & 3.92 \\ 
          \midrule
         Text Embedding &w/o input down& 1.55& 0.62 & 4.11 \\ 
        \midrule
         Backbone & w/o U-DiT (12.5 Hz) & 1.74&0.60&3.97 \\ 
        \midrule
         Objective & w/o GFM & 1.60& 0.59 & 3.98 \\ 
        \bottomrule
    \end{tabular}
    }\vspace{-0.1cm}
\end{table}
\vspace{-0.2cm}
\begin{table}[th!]
\caption{Analysis of NFE (Seed-en)}\vspace{-0.2cm}  \label{tab:tts_NFE}   
\centering
\resizebox{0.35\textwidth}{!}{
\begin{tabular}{l|ccc}
\toprule
Method  & \textbf{WER} $\downarrow$ & \textbf{SIM} $\uparrow$ & \textbf{UTMOS} $\uparrow$ \\
 \midrule
\multicolumn{4}{c}{\textbf{Different NFEs (Euler, Proportional Dur.)}} \\ 
 \midrule

4 NFE  &4.20& 0.67&3.05 \\
8 NFE  &1.93& 0.69&3.67 \\
10 NFE  & 1.85& 0.70&3.74 \\
16 NFE  & 1.77& 0.70&3.79 \\
25 NFE  & 1.62& 0.70&3.81 \\
32 NFE  & 1.67& 0.70&3.82\\ 
 \midrule
\multicolumn{4}{c}{\textbf{Different durations (25 NFE, GT Dur.)}} \\ 
 \midrule
1.5$\times$&9.28 & 0.71&3.65 \\
1.25$\times$&4.21 & 0.71&3.75\\
1.11$\times$&2.51 & 0.70&3.80 \\
1.0$\times$ (GT)& 1.68  & 0.70&3.81\\
0.9$\times$&1.71 & 0.69&3.80 \\
0.8$\times$ & 2.12 & 0.68&3.77 \\
0.67$\times$&6.01 & 0.66&3.69 \\
\bottomrule
\end{tabular}
}  \vspace{-0.5cm}
\end{table}

\vspace{-0.1cm}
\paragraph{Ablation Study} For ReGenTTS, we conducted an ablation study for efficient LDM-based TTS. With an average down-sampling-based implicit text-latent aligner, Table \ref{tab:tts_ablation_seed_benchmark} shows that the text encoder is not required for speech intelligibility. However, we adopt causal Transformers as a text encoder for better speaker similarity and word-level streaming generation. ConvNeXt blocks show under-fitted results compared to others. Due to highly-compress target latent, the initial downsampling layer for text embedding could enhance the efficiency and performance, and U-DiT structures including down/up-sampling layers for 6.25 Hz latent LDM improve the robustness of LDM training. Furthermore, the GFM objective improves the entire performance of LDM.

\vspace{-0.2cm}

\paragraph{Robust Speech Generation} We analyze the effect of NFE and duration control. Table \ref{tab:tts_NFE} shows the robustness of LDM even with a small number of NFE. While increasing the NFE over 25 can improve the audio quality, we chose 25 NFE for the trade-off between quality and speed. For a duration control, we use the GT duration, and the results show the robustness of duration control.

\vspace{-0.2cm}
\section{Conclusion} \vspace{-0.1cm} We propose representation generation for efficient waveform diffusion models via representation generation (ReGen). Compared to REPA, ReGen significantly improves the robustness and performance by improving the generative capacity. Furthermore, we introduce generalized flow matching (GFM) for better optimization of CFM by addressing the collapse of distinct stochastic paths into a single deterministic flow. With neural audio codec, VAE, and LDM-based TTS experiments, results verify the effectiveness of ReGen, GFM, and our proposed efficient waveform diffusion models.
\newpage
\vspace{-0.2cm}
\section*{Impact Statement}
\paragraph{Potential Broader Impact}
We propose novel generative models including ReGen and GFM. Then, we verify the effectiveness of ReGen and GFM by applying them to efficient waveform diffusion models for high-quality waveform generation. Specifically, neural audio codec and Wave-VAE can be utilized for speech generation and speech language models with better speech intelligibility and speaker similarity thanks to improved generative ability. Furthermore, LDM-based TTS models can be efficiently designed by operating latent diffusion at 6.25 Hz. This significantly reduce the inference time, leading to lower service cost. \vspace{-0.2cm}
\paragraph{Social Negative Impact} However, high-quality waveform generation models can be misused by cloning someone's voice to deceive others that can pose risks such as fraud and the spread of misinformation. Furthermore, the proposed LDM-based TTS models may be exploited to generate high similar synthetic speech without consent that can raise concerns regarding privacy issue. To mitigate these risks, it is essential to consider the watermarking or fake audio detection models together. Furthermore, we will release the source code with an explicit usage agreement and ethical guidelines to address misuse and promote responsible use.   
\vspace{-0.2cm}
\section*{Acknowledgments}\vspace{-0.2cm}
This work was supported by Institute of Information \& communications Technology Planning \& Evaluation (IITP) grant funded by the Korea government (MSIT) (RS-2025-02283048, Developing the Next-Generation General AI with Reliability, Ethics, and Adaptability, IITP-2026-RS-2023-00255968, the Artificial Intelligence Convergence Innovation Human Resources Development), National Research Foundation of Korea(NRF) grant funded by the Korea government (MSIT) (RS-2025-16069227), and the “Advanced GPU Utilization Support Program” funded by the Government of the Republic of Korea (Ministry of Science and ICT).
\vspace{-0.2cm}
\nocite{langley00}

\bibliography{regen}
\bibliographystyle{icml2026}

\newpage
\appendix
\onecolumn
\section{Implementation Details}
\begin{table*}[htb!]
  \caption{Hyperparameters of ReGen}
  \label{hyper}
  \centering
      \resizebox{0.9\textwidth}{!}{
  \begin{tabular}{c|l|c|c}
    \toprule
Module    &Hyperparameter          & ReGenTokenizer/ReGenVAE  & ReGenVoice-S/ReGenVoice  \\
\midrule
 Encoder  & Input Dim. & 1024 & 1024\\
 (Causal Transformer)& Hidden Dim. & 4096 & 4096 \\
 & Layer  & 8 & 6\\
 & Head & 16 & 16  \\
 \midrule
FSQ/VAE &  Hz & 25/12.5 & -  \\
 &  Dimension & 8([4,4,4,4,4,4,4,4])/32  &- \\
\midrule
Scale-DiT  & Input Dim. & 1024 & 1024/1152 \\
 & Hidden Dim. & 4096 & 4096/4608\\
 & Layer (U-net like)  & [3,3,6,3,3]& [4,10,4]/[6,12,6] \\
 & Head & 16 & 16/18  \\
 & AdaLN & AdaLN-SOLA (rank=32)& AdaLN-SOLA (rank=32)  \\
 \midrule
  Representation  & SSL & MMS (7th layers) & - \\
 \midrule
 Mel-spectrogram  & Hop & 480 & -\\
                  & Window & 1920 & -\\
                  & Bin & 256 & -\\
\midrule
Latent & Dimension & 32\\
 & Scaling & 0.1\\
 \midrule
Pre-train&Training Step& 1M & 1M \\
&Learning Rate& $1\times10^{-4}$ & $1\times10^{-4}$ \\
&Learning Scheduling& Warm-up (3,000 steps) & Warm-up (3,000 steps)  \\
&Batch Size& 256  & 256 \\
&Noise Scale (Wave, Mel, SSL) & 0.25, 1, 0.25 & 1\\
&Segment Size& 4s (96000 frames) & 2-30s\\
&Limited Context Attention Window& 4s & - \\
&Audio/Token Drop&0.3/0.2 &0.3/0.2\\
&$\lambda_{\text{regen}}$ & 0.1 &- \\
&$\lambda_{\text{neg}}$& 0.01 & 0.01 \\
&$\lambda_{\text{kl}}$& 1e-5 & -\\
& Model Size& 0.4B& 0.3/0.5B \\
\midrule
Adversarial Post-training&Training Step& 0.5M & -\\
&Learning Rate& $2\times10^{-5}$ & -\\
&Learning Scheduling& - & -  \\
&Batch Size& 128 & - \\
&Segment Size& 96000 & -\\
&Limited Context Attention Window& 96000 & -\\
&Audio/Token Drop&0.3/0 & -\\
& Weight of GAN Loss &1 & -\\
& Weight of Feature Matching Loss &2 & -\\
& Weight of STFT Loss & 0-1 (Warmup) & -\\
&$\lambda_{\text{kl}}$& 1e-2 & -\\
    \bottomrule
  \end{tabular}}
\end{table*}

We describe the hyper-parameters of ReGenTokenizer, ReGenVAE, and ReGenVoice in Table \ref{hyper}. For reproducibility, we will release all source code and checkpoints after paper notification. During pre-training of ReGenVAE, we regularize the latent by weak weight of KL loss(1e-5). Due to various losses for adversarial post-training, we utilize higher weight of KL loss (1e-2).
For ReGenVoice, we utilize the scaled latent representations by multiplying 0.1 to target latent representation of ReGenVAE. Then, before fed to the decoder of ReGenVAE, we re-scale the generated latent by multiplying 10. 

\newpage

\begin{table*}[t]
 \caption{Synthesis speed for each model. }  \label{table11:synthesis}
  \centering
      \resizebox{0.8\textwidth}{!}{
  \begin{tabular}{l|c|cccc|c}
    \toprule 
    Method  & Params. (M) & Hz & TPS & $N_q$&Stream.   &   xRT($\uparrow$)\\
    \midrule
SpeechTokenizer \cite{zhang2024speechtokenizer}& 103M &16k & 400 & 8 & \cmark &  $\times$76.899\\
\midrule 
BigCodec \cite{xin2024bigcodec} & 159M  &16k & 80 & 1  & \xmark&$\times$21.472  \\
X-codec2 \cite{ye2025llasa}& 822M &16k & 50 & 1& \xmark &$\times$14.026  \\
StableCodec \cite{parker2025scaling}& 953M &16k & 50 & 2& \xmark &$\times$63.456 \\
\midrule 
EnCodec \cite{dfossez2023high} & 15M &24k & 600 & 8 & \cmark &  $\times$64.267 \\
DAC \cite{kumar2023high}& 74M &24k & 600 & 8& \xmark &  $\times$60.237 \\
\midrule
Mimi \cite{defossez2024moshi}&  79M&24k & 100 & 8  & \cmark &$\times$45.538 \\
\midrule
WavTokenizer \cite{ji2025wavtokenizer}& 80M &24k & 75 & 1& \xmark &$\times$29.982\\
WavTokenizer \cite{ji2025wavtokenizer} & 80M &24k & 40 & 1 & \xmark & $\times$32.494\\
\midrule
\textbf{ReGenTokenizer}& 402M & 24k & 25 & 1 & \cmark & $\times$72.356\\
    \bottomrule
  \end{tabular}
  } 
\end{table*} 
\newpage

\section{Synthesis Speed}\label{appendix:speed}
We evaluate the synthesis speed relative to real-time (xRT) using an NVIDIA A100 GPU and report the model parameter size together. As shown in Table \ref{table11:synthesis}, our model achieves fast synthesis speed despite requiring four sampling steps for iterative generation. This efficiency is enabled by efficient waveform diffusion models (Wave-DiT) including linear-reshape transformation, diffusion at low-resolution (50 Hz), and AdaLN-SOLA.

\begin{table*}[th!]
\caption{Effect of prompted waveform generation}  \label{table:prompt_usage} \vspace{-0.2cm}
\centering
\resizebox{0.65\textwidth}{!}{
\begin{tabular}{l|ccc}
\toprule
Models&WER &SPK-SIM (to Prompt)  &SPK-SIM (to GT) \\
\midrule
\textbf{ReGenVAE-Emilia (w/o Prompt)}& 2.04&0.65& 0.86  \\
\textbf{ReGenVAE-Emilia (w Prompt)}& 2.09&0.73& 0.87  \\

\bottomrule
\end{tabular}
}  \vspace{-0.4cm}
\end{table*}

\section{Prompted Waveform Generation}
We evaluate the effectiveness of prompt for zero-shot cross-sentence style transfer in Table \ref{table:prompt_usage}. The results show that the latent of ReGenVAE already contains rich representations in terms of SPK-SIM (to GT). Furthermore, our new framework significantly improve the cross-sentence style transfer performance in terms of SPK-SIM (to Prompt).

\section{Limitation} While ReGen and GFM significantly improve the performance in low-bitrate and high-compress latent scenarios, our model still requires an adversarial post-training for better optimization and fast sampling speed. We see that there are still room for improvement to increase the capacity of waveform diffusion models without adversarial post-training. Recently, RAE \cite{zheng2025diffusion} adopts a wide decoupled diffusion Transformers (DDT) \cite{wang2025ddt} head for high-dimensional representation diffusion. We will apply the DDT head for waveform-level vector field estimation at low-resolution representation-level DiT. Furthermore, we have a plan to accelerate models via one-step generation using our hierarchical multi-prompting.

\section{Crowdsourcing Details} \label{apeendix:mturk} 
We conducted Mean Opinion Score (MOS) evaluations on a 5-point scale using crowdsourced listeners from Amazon Mechanical Turk. Each audio sample was independently rated by 20 native English speakers. MOS results in Table 2 were obtained from 208 utterances, and the ReGenVoice TTS evaluation used 300 samples randomly selected from the Seed-TTS dataset. To prevent unreliable scoring behavior, randomly inserted Gaussian noise samples were included as control items, and listeners who assigned high scores to these noise samples were excluded from the analysis. The total cost of all MOS evaluations was \$550.
\end{document}